\documentstyle[twoside,fleqn,espcrc2]{article}

\newcommand{\be}{\begin{equation}}
\newcommand{\ee}{\end{equation}}
\newcommand{\bea}{\begin{eqnarray}}
\newcommand{\eea}{\end{eqnarray}}
\newcommand{\lb}{\label}
\newcommand{\I}{\mbox{i}}
\newcommand{\D}{\mbox{d}}
\newcommand{\E}{\mbox{e}}
\newcommand{\AmS}{{\protect\the\textfont2
  A\kern-.1667em\lower.5ex\hbox{M}\kern-.125emS}}

\title{Origin of classical structure from inflation}

\author{Claus Kiefer\address{Faculty for Physics, 
        University of Freiburg, \\ 
        Hermann-Herder-Strasse~3, 79104 Freiburg, Germany}}

\begin{document}

\begin{abstract}
According to the inflationary scenario, all structure in the Universe
can be traced back to quantum fluctuations of the metric and scalar field(s)
during inflation. The seeds of this structure can be observed as
classical anisotropies in the cosmic microwave background.
 I briefly review how the transition from the
inherent quantum nature of these fluctuations to classical behaviour
comes about. Two features play a crucial role: Firstly, the quantum state
of the fluctuations becomes highly squeezed for wavelengths that
exceed the Hubble radius. Secondly, decoherence due to other fields
distinguishes the field-amplitude basis as the classical pointer basis.
I also discuss the entropy of the fluctuations and make
a brief comparison with chaotic systems.
\end{abstract}

\maketitle
The assumption that the Universe underwent an accelerated
(``inflationary'') expansion during its early phase provides
a theoretically satisfying scenario for a causal explanation
of structure formation \cite{KT}. How does this work?

It is evident that all pre-existing structure
would have been damped away during the rapid
inflationary expansion. The only exception are
the ubiquitous {\em quantum fluctuations} 
of fundamental fields. In fact, the central idea
is that all observed structure in the Universe
(galaxies, clusters of galaxies, etc.) has
its origin in quantum fluctuations that are present during
inflation. These fluctuations can be observed
as classical stochastic fluctuations in the
anisotropy spectrum of the cosmic microwave
background radiation (CMB). Such observations
are made both from space through satellite missions
such as COBE (and the planned missions PLANCK and MAP)
 as well as from earth-bound telescopes.
The main issue to be addressed here
is: How can the transition from quantum to
classical fluctuations be understood?
The analysis will consist of two parts: Firstly, the
dynamics of the fluctuations as an isolated system
is discussed. Secondly, their interaction with
other fields (``environment'') is taken into account.

The simplest example, which nevertheless captures the
essential physical features, is the model of a real
massless (minimally coupled) scalar field $\phi$
in a flat Friedmann Universe with scale factor
$a$. It readily applies
to the case of gravitational waves and can be extended
in a straightforward way to scalar perturbations by
employing a gauge-invariant formalism \cite{BHM,Al94}.
The action reads
\be
S=\frac{1}{2} \int\ \D^4x
 \sqrt{-g} ~\partial^{\mu}\phi ~\partial_{\mu}\phi~.
\ee
It turns out convenient to introduce the rescaled variable $y\equiv a\phi$
(the corresponding momenta thus being related by $p_y=a^{-1}p_{\phi}$).
Assuming for simplicity that the volume of space is finite, the
corresponding Hamiltonian can be written as a sum 
\be H=\sum_{\bf k} H_{\bf k}~, \ee
where
\be H_{\bf k}= \frac{1}{2}\left(p_{\bf k} p_{\bf k}^* + 
k^2y_{\bf k}y_{\bf k}^* +
    \frac{a'}{a}\left[y_{\bf k}p_{\bf k}^* + p_{\bf k}y_{\bf k}^*
\right]\right)  \lb{ham} \ee
and $p_{\bf k}$ is given by 
$p_{\bf k}=y_{\bf k}'-\frac{a'}{a} y_{\bf k}$; it is the Fourier transform 
of $p_y=y'-\frac{a'}{a} y$, the momentum conjugate to $y$.
Primes refer to derivatives with respect to conformal time
$\eta\equiv \int  dt/a$. All quantities are evaluated
in Fourier space. The physical wavelength of a perturbation
is $\lambda_{phys}=a/k$.
 It is well known from quantum optics
that a Hamiltonian of this type generates a 
{\em two-mode squeezed state}. The application of this formalism
to cosmological fluctuations was discussed extensively in
\cite{Al94,GS,PS}.

What is the initial quantum state for the perturbations?
A natural assumption would certainly be to take the
vacuum state for the fluctuations
 at the onset of inflation. This will be done
here, but it must be emphasised that the results are insensitive
to the exact choice of initial condition: Inflation leads to 
a very high squeezing in momentum and, unless the initial
state is ``fine-tuned'' to be extremely squeezed in position,
all differences in the initial state are damped out by inflation.
It is of course a deep and non-trivial question why inflation
does occur in the first place.
This seems to be related with the origin of irreversibility
in the Universe, a question that can be dealt with
satisfactorily only in the framework of quantum
gravity \cite{zeh}.   

The squeezed vacuum can be written,
in the Schr\"odinger picture, as a Gaussian wave function
for each mode \cite{KP},
\be
\psi_{k0}=\left(\frac{2\Omega_R}{\pi}\right)^{1/2}
        \exp\left(-[\Omega_R+\I\Omega_I]|y_k|^2\right)~.  \lb{psik0}
\ee
Since modes with different wave vectors ${\bf k}$ decouple,
I shall sometimes skip the index ${\bf k}$ (or $k$) in the following.
The wave function (\ref{psik0}) can be written in terms
of the squeezing parameters $r$ and $\varphi$ in the form
\bea
\psi_0&=&\left(\frac{2k}{\pi[\cosh2r+ \cos2\varphi\sinh2r]}\right)^{1/2}
 \nonumber\\
 & & \times \exp\left(-k\frac{1-\E^{2\I\varphi}\tanh r}
                  {1+\E^{2\I\varphi}\tanh r}|y|^2\right)\ . \lb{psi0}
\eea 
Squeezing can equivalently be expressed in terms of
``particle creation'' with average particle number $N({\bf k})\approx
\E^{2r}/4\approx aH_I/4k$, where $H_I$ denotes the Hubble parameter
of inflation, and the limit of large squeezing $r\to\infty$ is taken.
This corresponds to modes with wavelengths bigger than the horizon
(after the first Hubble-radius crossing during inflation).
Due to this large squeezing (for the largest cosmological scales
one has $r_k\geq 100$), the system already exhibits a certain
degree of classicality by itself, called ``decoherence without
decoherence'' in \cite{PS}. In the Heisenberg picture, this is
recognised as the negligibility of the decaying mode:
If terms proprotional to $\E^{-r}$ can be neglected, position
and momentum approximately commute. Thought experiments and the
analogy with a free quantum particle can be invoked
to illustrate this effective classicality \cite{KP,KLPS,Pol}. 
This classicality is of a stochastic nature and different
from, for example, the classicality associated with coherent-state
evolution. Due to this large squeezing (which one can recognise
directly from the Wigner ellipse in phase space), the system
exhibits many similarities with a chaotic system, although it 
is not chaotic, but classically unstable. This will be of relevance
for an understanding of the entropy, see below. It must
also be emphasised that these properties (random amplitude
and fixed phase) are preserved for a sufficient time
after the modes re-enter the horizon in the 
post-inflationary phase \cite{KLPS} and lead to 
a distinctive signature of inflationary models in the form
of peaks exhibited by the CMB-anisotropy spectrum.

Classical properties arise from quantum systems usually
through the ubiquitous interaction with environmental degrees
of freedom -- the irreversible process of decoherence
\cite{deco}. Although, as discussed above, the system exhibits
classical features by itself, the interaction with other
degrees of freedom is unavoidable, and it is important to
check that this influence does not spoil the properties of the
isolated system. Since the relevant perturbation modes
are outside the horizon during inflation, direct causal
processes cannot occur and relaxation is therefore
negligible. However, quantum entanglement can and does
indeed occur \cite{KP}. Since all relevant interaction
of the perturbations with other degrees of freedom
couple field variables (as opposed to the corresponding
momenta), the classical pointer basis is given by
the field amplitudes $|y_{\bf k}\rangle$. The corresponding
operator thus commutes with the interaction Hamiltonian, 
\be
[H_{int},y_{\bf k}]\approx 0\ .
\ee
Moreover, in the large-squeezing limit, the field amplitudes
at different times commute,
\be
[y_{\bf k}(t_1),y_{\bf k}(t_2)]\approx 0\ ,
\ee
and therefore constitute quantum-nondemolition variables
\cite{KPS1}.
Of course, the last two equations together imply that
$y_{\bf k}$ commutes with the full Hamiltonian, which is therefore
an approximate constant of motion. 
The easiest interaction with an environment, giving a
lower bound on decoherence, is a linear coupling with other
fields like in quantum Brownian motion \cite{deco}.
In such a model one finds for the decoherence time \cite{KP}
\be
t_D\approx \frac{\lambda_{phys}}{g\E^r}\approx (gH_I)^{-1}\ ,
\ee
where $g$ denotes the (dimensionless) coupling constant.
If one assumes that $g$ is of order one, one has 
$t_D\approx H_I^{-1}$, i.e., the decoherence timescale
is set by the Hubble parameter of inflation
and independent of the details of the interaction!
After the perturbation has remained outside the horizon
for a couple of e-folds, the $|y_{\bf k}\rangle$ become
a perfectly classical pointer basis. For this reason, the
results drawn from the discussion of the system itself
remain unchanged, which would not be true if, for example,
the particle-number basis would be the classical basis.
The interaction constitutes an ``ideal measurement'':
All interferences disappear from the system, while all
probabilities (predictions of the inflationary scenario)
remain unchanged, see \cite{KP} for a detailed conceptual
discussion. 

The above results are of direct relevance for the
calculation of the entropy possessed by the pertubations
\cite{KPS2}. In the following I shall focus on the gravitons,
but a similar discussion can be made for the matter
perturbations. Clearly, for the isolated system the entropy
is and remains zero, since the system is in a pure state
\cite{LPS}. A non-vanishing entropy arises from the interaction
with the environment, since then the perturbations
become an open system to which one can only attribute
a (reduced) density matrix \cite{deco}. The maximal entropy
for each mode is obtained if the squeezed ellipse in phase space
is coarse-grained to a big circle, yielding 
an entropy $S_k\approx 2r_k$. This would mean that the 
squeezing phase would be maximally randomised and that
therefore some features which would otherwise be present in the
CMB anisotropy spectrum would be absent. 
As the discussion in \cite{KPS2} demonstrates,
the actual entropy is much smaller than its maximum value
and lies between zero and $2r_k$. The point is that 
a few bits of information loss are sufficient for
decoherence, and classicality does already emerge if
the entropy is (in a logarithmic sense) much bigger
than about $1-\ln 2\approx 0.31$.

 The entropy is calculated
via the von Neumann formula $S=-\mbox{Tr}(\rho\ln\rho)$,
where $\rho$ is the reduced density matrix, and use is being made
of explicit formulae derived in \cite{JZ}. The derivation
confirms the expectation that the entropy is roughly given by the
logarithm of the ellipse in phase space, whose minor axis
approaches a finite value due to decoherence (as opposed
to ongoing squeezing). 

In addition to environmental decoherence, there might also be
some information loss due to the presence of a
secondary gravitational-wave background caused by
postinflationary processes, since the corresponding
gravitons are indistinguishable from the
gravitons belonging to the primordial background \cite{KPS2}.
 The result for the entropy
depends on the corresponding particle numbers, and various
cases are discussed in \cite{KPS2}.
The general expectation is that the secondary background is
much smaller than the primary one,
 so that the squeezing is still recognisable.

It turns out that, after a complicated transitory phase,
the entropy production rate is given by the Hubble
parameter, i.e. $\dot{S}\approx H$, in accordance with
the result that the decoherence time is proportional
to $H^{-1}$. During inflation, $H=H_I\approx const.$,
and the entropy production is linear in $t$. 
In the post-inflationary phases, the Hubble parameter is
proportional to $1/t$, and the entropy increases
only logarithmically in time.

It is of interest that the situation is very similar to
chaotic systems \cite{PZ}. The role of the Hubble parameter
is there played by the Lyapunov exponent $\lambda$.
One can define a measure of entropy production, the
so-called Kolmogorov entropy, see e.g. \cite{Sch}, 
which measures how chaotic a dynamical system is, i.e.
how fast information about the state is lost. The Kolmogorov
entropy corresponds to the entropy production referred to
above. The reason for this analogy is the fact that
the situation in phase space is similar in
both cases.

\vskip 5mm

It is a pleasure to thank David~Polarski and Alexei~Starobinsky
for their collaboration during which
the results reported in this contribution have been obtained.
I am also grateful to David Polarski
and Sebastian Schlicht for a careful reading of this
manuscript.

\end{document}